\documentclass[]{spie}  

\usepackage{amsmath,amsfonts,amssymb}
\usepackage{graphicx}
\usepackage{caption}
\usepackage{subcaption}
\usepackage[colorlinks=true, allcolors=blue]{hyperref}
\usepackage{multirow}
\usepackage[perpage]{footmisc}

\title{RNDR noise modeling in first-generation Single electron Sensitive Readout (SiSeRO) devices}

\author[a]{Tonya L. Peshel}
\author[a,b]{Abigail Y. Pan}
\author[a]{Tanmoy Chattopadhyay}
\author[a,b,c]{Steven W. Allen}
\author[d]{Marshall W. Bautz}
\author[e]{Michael Cooper}
\author[e]{Kevan Donlon}
\author[d]{Catherine E. Grant}
\author[a]{Sven Herrmann}
\author[d]{Jill Juneau}
\author[d]{Beverly J. LaMarr}
\author[e]{Christopher Leitz}
\author[a]{Adam B. Mantz}
\author[d]{Eric D. Miller}
\author[a,c]{R. Glenn Morris}
\author[a]{Declan O'Neill}
\author[a]{Peter Orel}
\author[a]{Artem Poliszczuk}
\author[d]{Gregory Y. Prigozhin}
\author[a,b]{Haley R. Stueber}
\author[e]{Keith Warner}

\affil[a]{Kavli Institute for Particle Astrophysics and Cosmology, Stanford University, 452 Lomita Mall, Stanford, CA 94305, USA}
\affil[b]{Department of Physics, Stanford University, 382 Via Pueblo Mall, Stanford CA 94305, USA}
\affil[c]{SLAC National Accelerator Laboratory, 2575 Sand Hill Road, Menlo Park, CA 94025, USA}
\affil[d]{MIT Kavli Institute for Astrophysics and Space Research, Massachusets Institute of Technology, 70 Vassar St, Cambridge, MA 02139, USA}
\affil[e]{MIT Lincoln Laboratory, 244 Wood St building 1324, Lexington, MA 02421, USA}

\authorinfo{Further author information: Send correspondence to T.L. Peshel,  tpeshel@stanford.edu\\}

\begin{document} 
\maketitle


\begin{abstract}
Flagship observatories require single-photon detectors with ultra-fast readout, sub-electron noise performance, and scalable large-format architectures. The X-ray Astronomy and Observational Cosmology group at Stanford, in collaboration with the MIT Kavli Institute and MIT Lincoln Laboratory, is developing readout technologies for next-generation detectors. Prototypes employing Single-electron Sensitive Readout (SiSeRO) amplifiers demonstrate excellent read noise and spectral performance using repetitive non-destructive readout (RNDR), achieving 0.5 e$^-$ noise in under 57 cycles. We have modeled noise for longer RNDR cycles, exploring probabilistic mechanisms such as thermal leakage and impact ionization. Here we present our model results, including statistical limits on dark current-like signals. Maturation of SiSeRO technology will improve detector performance at soft X-ray energies,  addressing technology gaps for future X-ray and UV/visible/near-IR observatories.
\end{abstract}



\section{INTRODUCTION}
\label{sec:intro}  

Future flagship observatories operating in the X-ray and UV/visible/near-IR regimes will require substantial advances in detector technology beyond the capabilities of conventional charge-coupled devices (CCDs). The Astro2020 Decadal Survey report identified advanced detector development as a critical enabling technology for next-generation missions, emphasizing the need for sensors capable of sub-electron noise performance, ultra-fast readout rates, and scalability to large focal plane arrays\cite{nationalacademiesofsciencesPathwaysDiscoveryAstronomy2023}. For example, the Habitable Worlds Observatory (HWO) will be designed to directly image Earth-like exoplanets at $<$100 mas spatial resolution. To achieve this sensitivity, CCD detectors for the HWO coronograph system will require single-counting photon ability with $<$0.1 e$^-$ read noise and dark current below 10$^{-4}$ e$^-$/pixel/second in a 4k$\times$4k pixel format\cite{HabWorldsRoadmap2024}. Similarly, future large ground-based telescopes will require detectors capable of high frame rates and very low read noise. 

To address this technology gap, the X-ray Observational Astrophysics and Cosmology (XOC) group at Stanford University, in collaboration with MIT Lincoln Laboratory (MIT-LL) and the MIT Kavli Institute for Astrophysics and Space Research, is developing novel detector and readout technologies for next-generation spectro-imagers\cite{bautz20,herrmann20_mcrc,Bautzetal2022,chattopadhyay22_ccd,bautz_ccd_2024,porelMCRCspie2024,stueber_ccd_2025,herrmann2024}. A newly developed device architecture known as Single-electron Sensitive Readout (SiSeRO\cite{chattopadhyay22_sisero,chattopadhyayImprovedNoisePerformance2023}) presents a promising path toward extending CCD technology into the ultra-low-noise regime. 

\begin{figure}[htb]
    \centering
   \includegraphics[width=.50\textwidth]{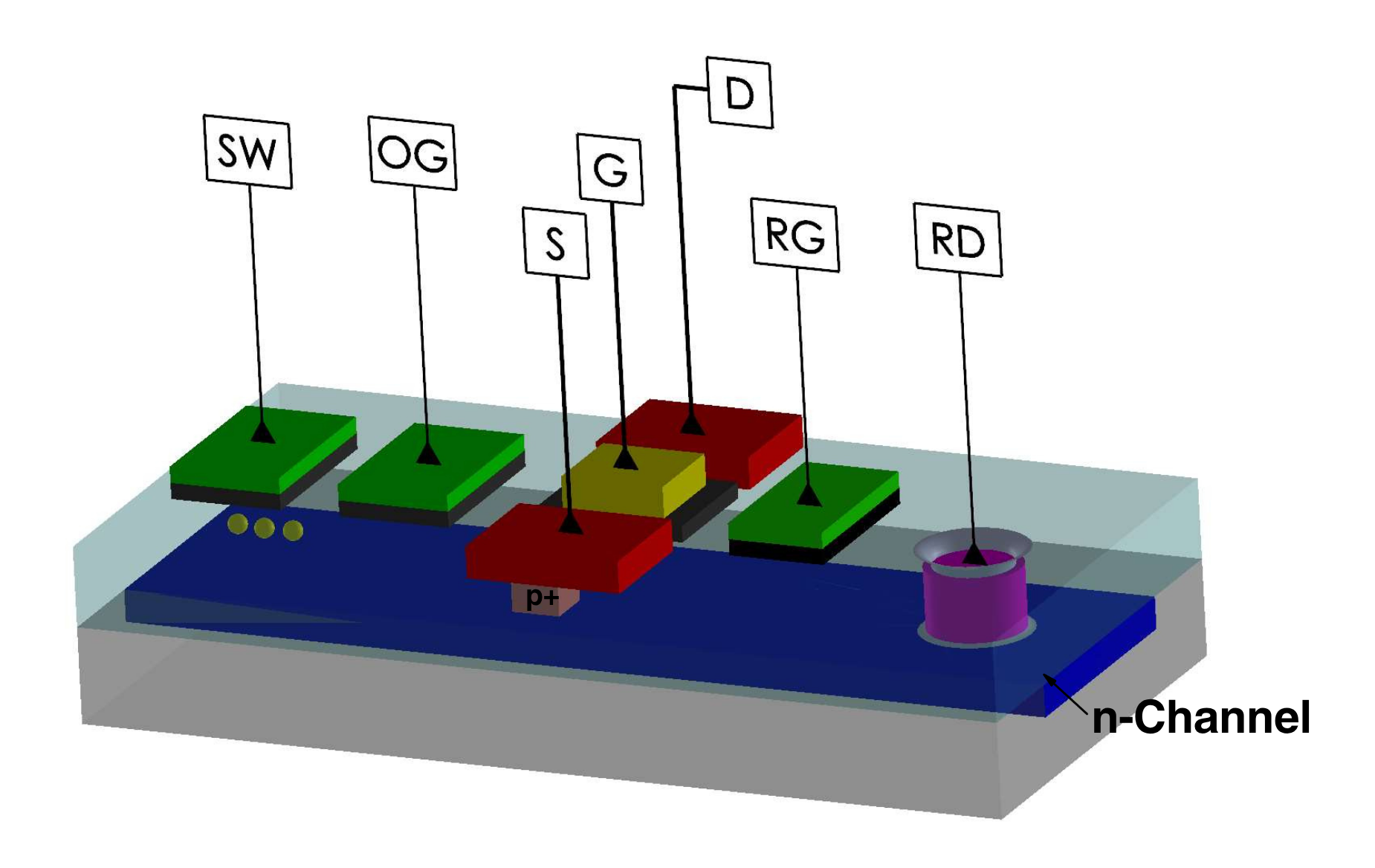}
   \includegraphics[width=.48\textwidth]{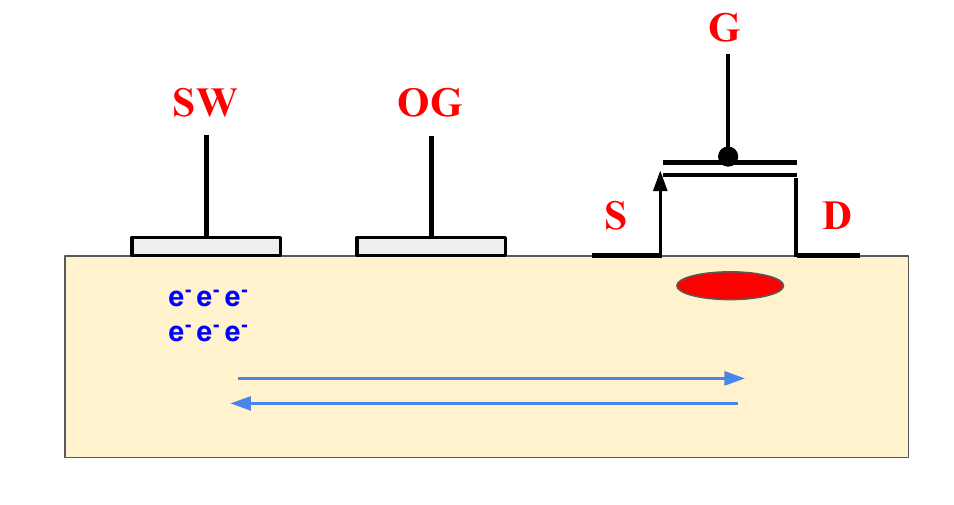}
    \caption{Left: Schematic of the buried-channel SiSeRO output stage\cite{chattopadhyay22_sisero}. Charge is transferred from the last serial gate (summing well, SW) across the output gate (OG) to a p-MOSFET transistor (source S, drain  D, and gate G terminals). The transistor sits above an internal gate, which is emptied by the reset gate (RG) and reset drain (RD). The transistor drain current is modulated when a charge packet is transferred from the SW and OG into the internal gate. Right: In repetitive non-destructive readout (RNDR), the signal charge is transferred repeatedly between the summing well and the internal gate to make multiple independent measurements\cite{chattopadhyay24_rnrdr}. }
    \label{fig:sisero_diagram}
\end{figure}

First-generation CCID-93 SiSeRO prototypes, developed by MIT-LL, contain a buried-channel p-MOSFET SiSeRO amplifier at the output stage (Fig. \ref{fig:sisero_diagram} left). They consist of a 4$\times$4 mm imaging area with a 512$\times$512 pixel array of 8 $\mu$$m$ pixels. The prototypes are tested in our X-ray beamline\cite{stueber2024,panDesignDevelopmentCommissioning2025} at a temperature of $-100$$^{\circ}$ C (173 K). A detailed description of the detector design can be found in Chattopadhyay et al 2022\cite{chattopadhyay22_sisero}. 

A key advantage of the SiSeRO detectors is their ability to perform repetitive non-destructive readout (RNDR). In RNDR, a single charge packet is repeatedly shuffled in and out of the internal gate of the SiSeRO output stage non-destructively (Fig. {\ref{fig:sisero_diagram}} right). The charge packet remains preserved, allowing for many independent measurements of the signal. Averaging these measurements reduces the effective read noise proportional to $1/\mathrm{\sqrt{N}}$, where N is the number of RNDR cycles. The RNDR technique has been demonstrated to provide sub-electron noise yield in Skipper CCDs \cite{tiffenberg17,rodrigues21,Lapi2024_skipperMAS}, DEPFET devices \cite{wolfel06,treberspurg22_rndrdepfet}, and in a prototype CMOS imaging sensor (CIS) \cite{Stefanov2020_skippercmos,Lapi2024_skipperCMOS}.

First-generation SiSeROs have already demonstrated exceptional read noise and spectral performance at fast speeds. The best demonstrated performance to date achieves a spectral resolution of 132 eV full width at half maximum (FWHM) at the 5.9 keV Mn K$\alpha$ line, 3.8 $e^-$$_{RMS}$ for the first RNDR cycle at a speed of 625 KHz, single electron read noise after only 15 RNDR cycles, and 0.49 $e^-$$_{RMS}$ noise after 57 cycles at an effective readout speed of 10 KHz\cite{chattopadhyayDemonstratingSubelectronNoise2024, PanJATIS2026}. These results establish SiSeRO technology as a promising detector candidate for the future of observational astronomy.  

Notably, the cumulative noise for our longest RNDR  measurements (up to 200 cycles) do not deviate from the theoretical $1/\mathrm{\sqrt{N}}$ trend\cite{PanJATIS2026}. However, as we increase the number of RNDR cycles to push toward lower noise requirements, we may eventually see a divergence due to different random probabilistic phenomena. Such stochastic mechanisms may introduce excess dark current-like noise that scales linearly with the pixel read time, limiting the final noise performance. 

To simulate detector behavior over more RNDR cycles and investigate the limits at which we will see this excess noise, we have developed a Monte Carlo based noise model for the SiSeRO output stage operating in RNDR mode. The model is generic and applicable to any RNDR device. Here, we describe the main components of the model and summarize its predictions for two potential noise sources: thermal charge leakage and impact ionization.


\section{Modeling Dark Current-Like Noise Sources} 
\label{sec:modeling}  

The Monte Carlo-based noise model we developed to simulate the noise behavior for large RNDR cycles is based on the first-generation SiSeRO architecture. The model includes transistor thermal noise, trap-induced flicker ($1/f$) noise, and excess dark current-like noise arising from trap-induced thermal leakage and impact ionization.


\subsection{Generic noise model for RNDR devices} \label{sec:noise_model}

To construct our model, we first generate a digitized waveform (matching the 10 ns resolution of the Archon controller\cite{bredthauerArchonModernController2014} ADC sampling rate) consisting of a baseline, a signal region, and a reset period. The length of each cycle matches the correlated double sampling period of the actual data (1.6 $\mu$$s$). Next, we simulate the RNDR process by subtracting the average signal value from the average baseline for each RNDR cycle and multiplying by the gain, resulting in an array of the individual CDS values for each cycle as well as each pixel. To calculate the noise of the individual cycles, we take the standard deviation of charge across all pixels for a given cycle. To calculate the cumulative noise at the N$\mathrm{^{th}}$ cycle, we take an average across all N cycles in a given pixel, then find the standard deviation across all averaged pixels. In an ideal model with no noise, the cumulative noise is consistent with the $1/\mathrm{\sqrt N}$ trend. 

We first modeled the read noise of the output stage using a thermal noise component and a $1/f$ (trap-induced flicker) noise component in the SiSeRO MOSFET amplifier. The remainder of the readout electronics contribute negligible noise and are therefore ignored in the model\cite{porelMCRCspie2024}. The primary contribution to the output stage thermal noise comes from the resistive channel of the MOSFET, with current noise density proportional to the square root of temperature (T) and transconductance of the channel \cite{van_fet_noise_1962}. For simplicity, we injected the RNDR waveform with a random scatter drawn from a Poisson distribution around the known read noise of the detector at 173K in single read mode. To account for temperature dependence, we scale the mean read noise as $\mathrm{\sqrt{T/T_{173K}}}$. The $1/f$ noise is modeled following a prescription given in Pullia et al. 2004\cite{Pullia04} using a Poissonian sequence of pulses in the time domain with a fixed amplitude and random sign.
 
Because the focus in this paper is the noise of the output stage, we do not model thermal dark current or other noise features expected to originate in the image region of the detector. 

\begin{figure} [h!]
\centering
    {\includegraphics[width=0.5\textwidth]{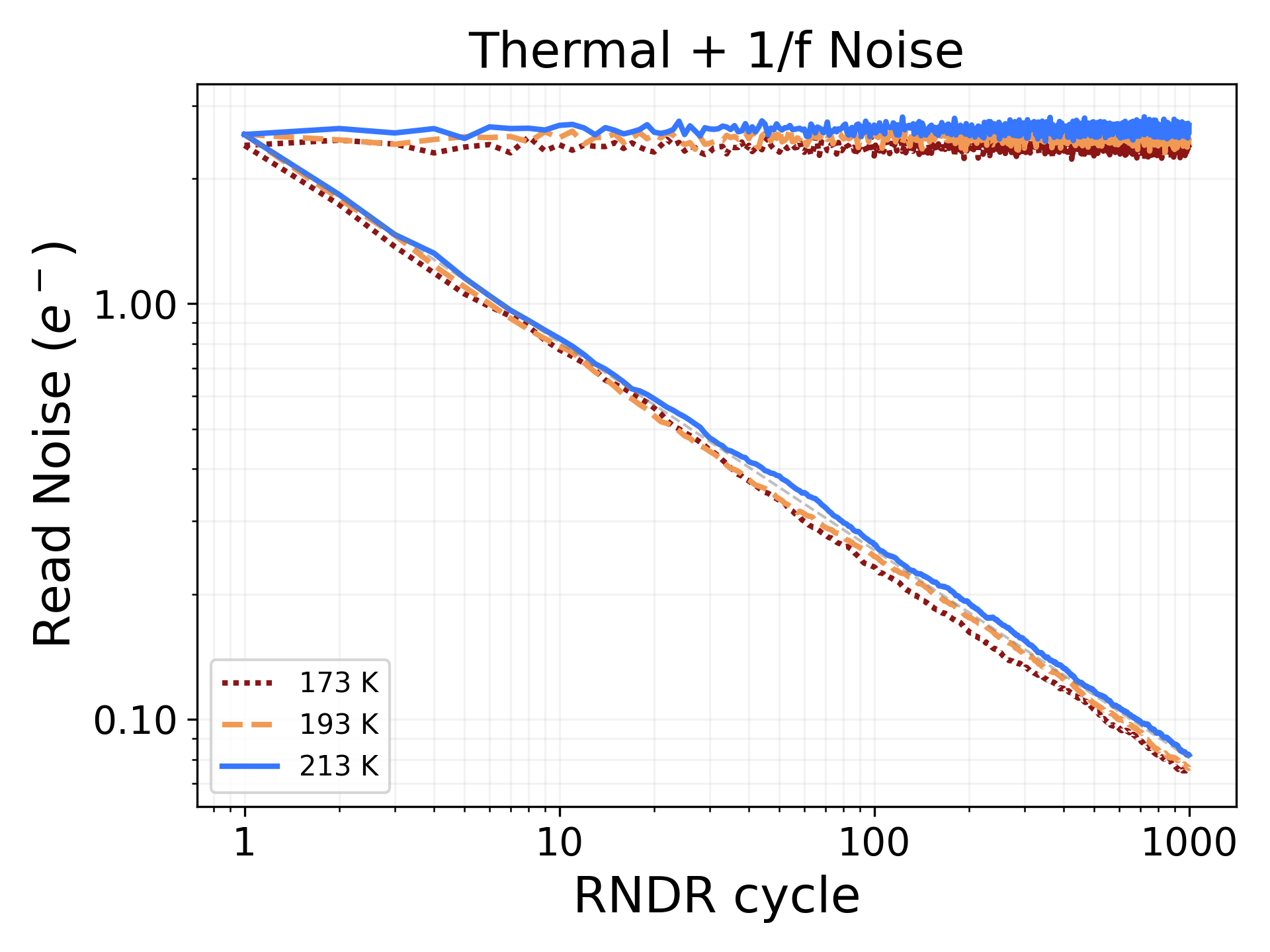}}
\caption{Model of the SiSeRO RNDR mode noise with thermal and $1/f$ noise components.}
\label{fig:readnoise_and_1f}  
\end{figure}

Adding $1/f$ noise and thermal read noise into our model did not introduce any divergence from the $1/\mathrm{\sqrt N}$ trend, either for large RNDR cycles or any arbitrarily large values of read noise and $1/f$ noise (Fig. \ref{fig:readnoise_and_1f}). This behavior matches our expectations of an ideal RNDR readout device in the absence of any time-dependent excess noise. 

We next explored two possible processes that can introduce dark current-like excess noise: thermally generated charge carriers in the depleted internal gate region of the output stage and impact ionization in the SiSeRO MOSFET transistor channel.


\subsection{Trap-Assisted Thermal Leakage} \label{sec:thermal_leakage}

\begin{figure} [t!]
\centering
\begin{subfigure}{0.32\textwidth}    {\includegraphics[width=\textwidth]{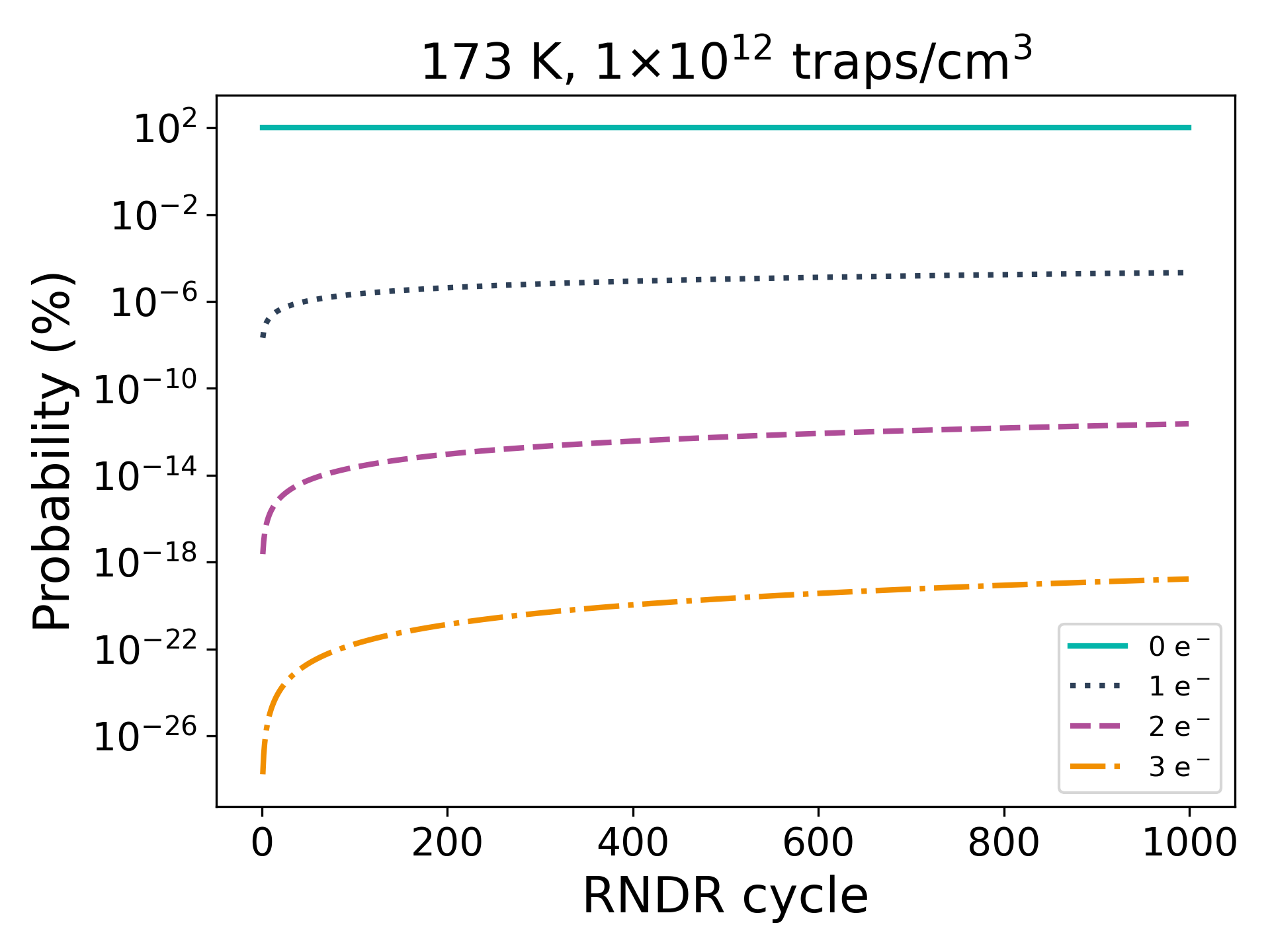}}
\end{subfigure}
\begin{subfigure}{0.32\textwidth}
    \includegraphics[width=\textwidth]{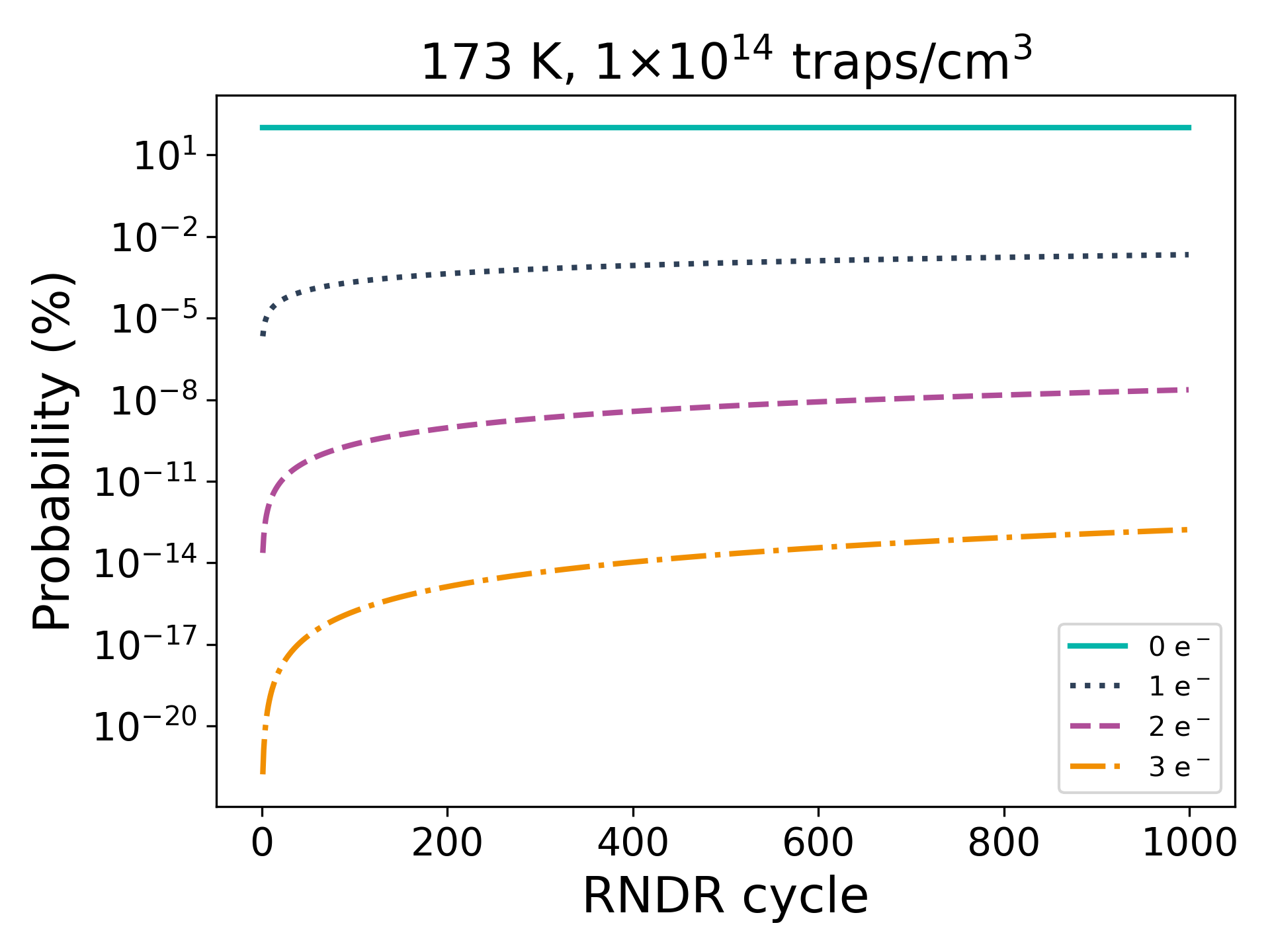}
\end{subfigure}
\begin{subfigure}{0.32\textwidth}
    \includegraphics[width=\textwidth]{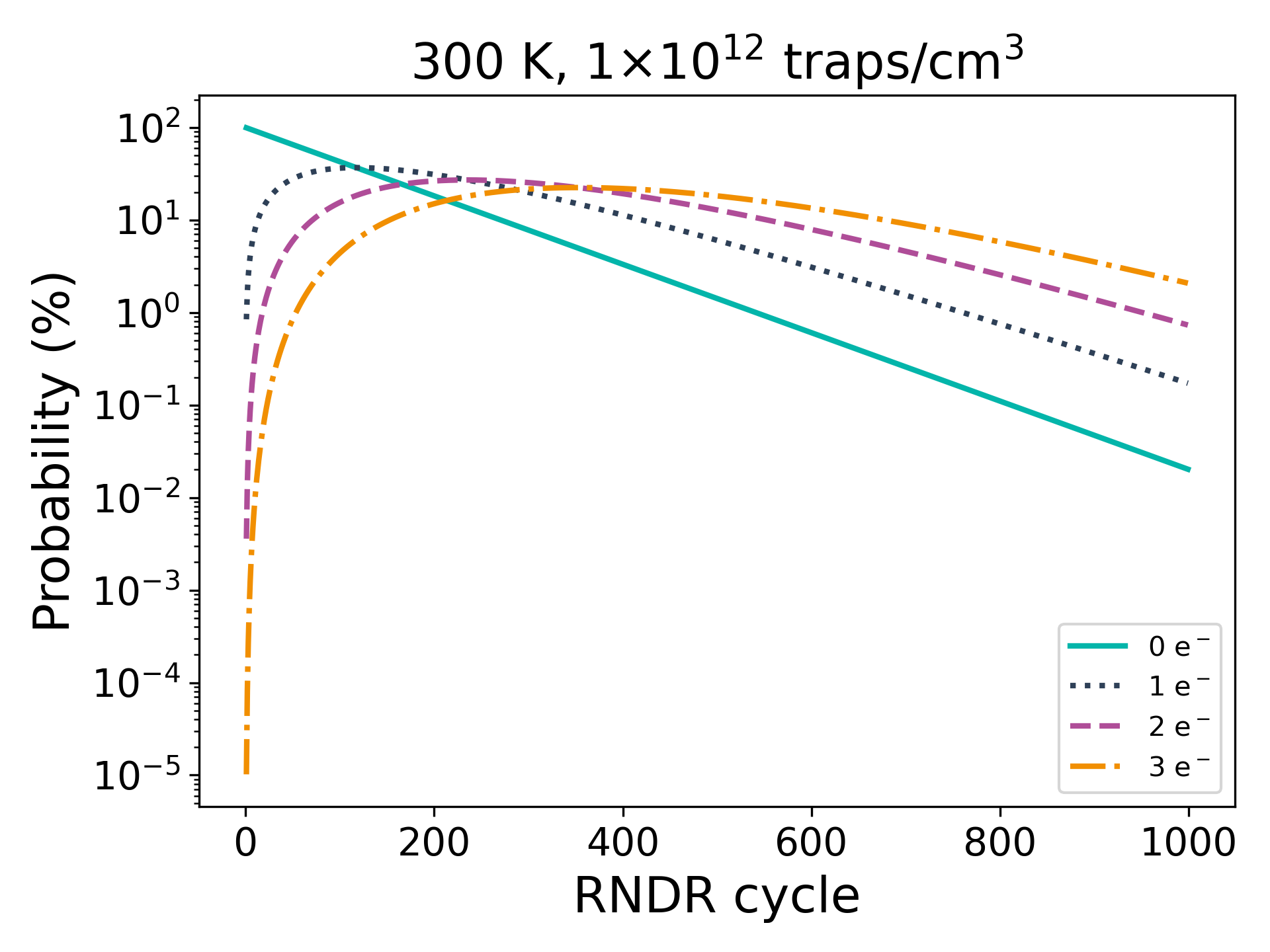}
\end{subfigure}
\caption{Probability of generating 0, 1, 2 or 3 electrons via trap-assisted thermal generation. The probability increases with trap density and with temperature.}
\label{fig:thermal_prob_percent_model}  
\end{figure} 

Crystal defects in the silicon introduce intermediate energy levels in the silicon band gap and thereby assist in the generation of electron-hole pairs in the SiSeRO depleted internal gate region. With more RNDR cycles (therefore longer integration times), we anticipate a higher probability of thermal leakage. Generation of electron-hole pairs in the depletion region is modeled using the Shockley-Read-Hall (SRH) model \cite{Shockley52,Sah57}, which depends on the effective density of states in the conduction band, the energy and density of the defect states in silicon, the trapping cross-section, and the physical temperature of the device.

We first calculate the rate of electron emission for a given temperature per trap. The rate of emission is 
\begin{equation}
\label{eq:r_emit_thermal}
R_{emit}=(\sigma v_{th}n_c)exp[-(E_C-E_t)/kT] \, ,
\end{equation}
where $\sigma$ is the capture cross section of the trap, $v_{th}$ is the thermal velocity of the charge carriers, $n_c$ is the effective density of the states in the conduction band, $E_c$ is the energy of the conduction band edge, and $E_t$ is the trap energy state.  Since the process is most effective for mid-level traps, we keep the trap energy constant at $\sim$0.6 eV for the calculations. 

\begin{figure} [hb]
\centering
    {\includegraphics[width=0.5\textwidth]{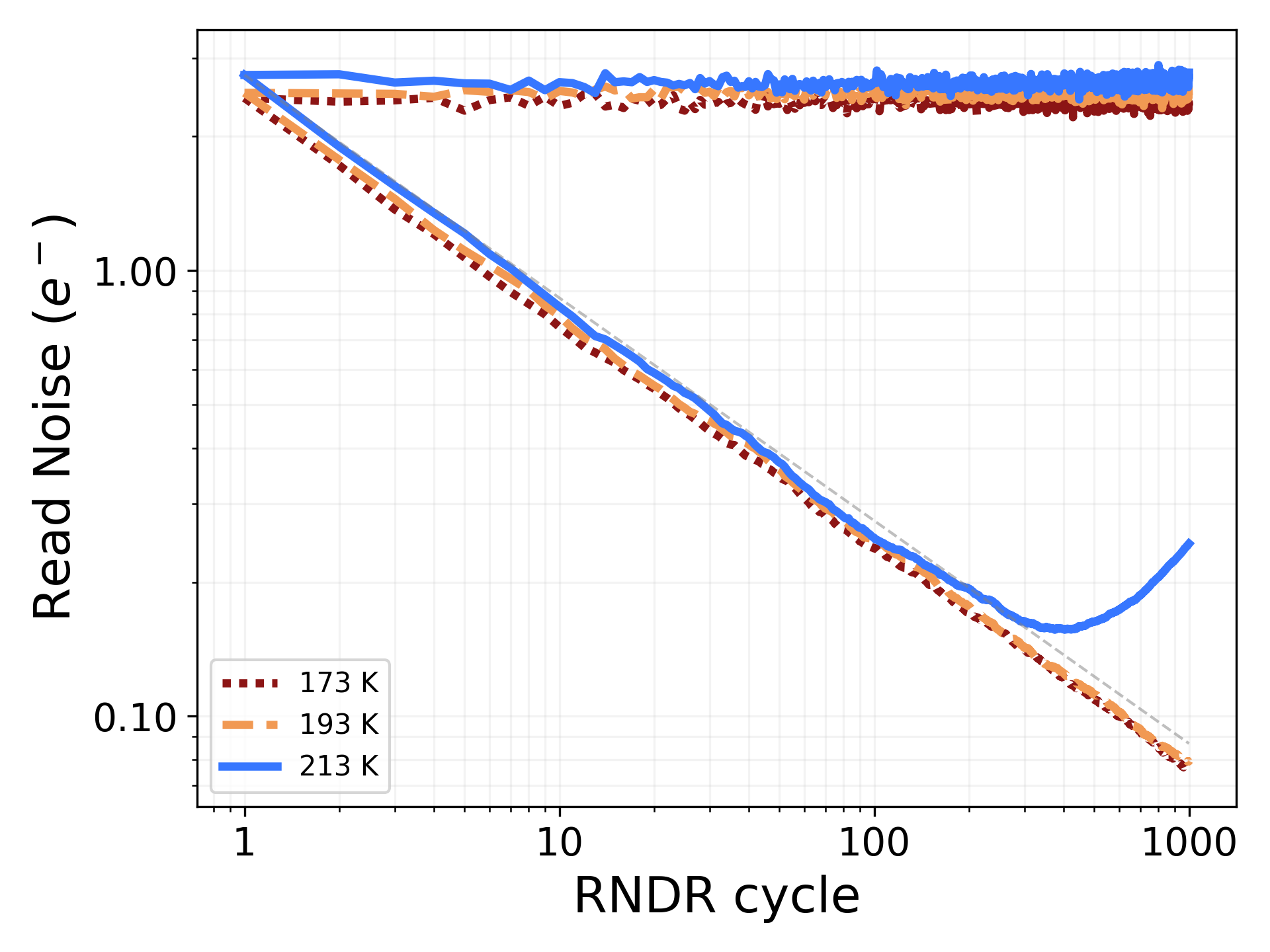}}
\caption{Simulation of the effect of trap-assisted thermal leakage across 1000 RNDR cycles for three temperatures for a trap concentration of $1\times10^{12}$ traps per cm$^3$. The model includes a hypothetical near-maximal level of trap-assisted thermal leakage that remains undetected in the current laboratory data\cite{PanJATIS2026}. The model predicts no thermal leakage effects at our operating temperature of 173 K.  
}
\label{fig:thermal_leakage_e}  
\end{figure}
The total number of electrons emitted in time $t$ is 
\begin{equation}
\label{eq:N_emit_thermal}
N_{emit}= R_{emit}N_{trap}V_{dep}~t \, ,
\end{equation}
where $N_{trap}$ is the trap density and $V_{dep}$ is the volume of the output stage.

Using Poisson statistics, we then calculate the probability of emission of $n$ electrons:

\begin{equation}
\label{eq:prob_thermal_emission}
P(n)= \frac{N_{emit}^n exp(-N_{emit})}{n!}\, 
\end{equation}

We find that the probability increases with temperature and trap density (Fig. \ref{fig:thermal_prob_percent_model}). This probability is then incorporated into our Monte Carlo model to calculate the read noise as a function of RNDR cycles. The model predicts negligible contribution of thermal leakage noise at our operating temperature of 173 K (Fig. \ref{fig:thermal_leakage_e}). The excess noise becomes significant only at higher operating temperatures.


\subsection{Impact Ionization} 
\label{sec:impact_ionization}

\begin{figure} [h!]
\centering 
    \raisebox{-10mm}
    {\includegraphics[width=0.5\textwidth]{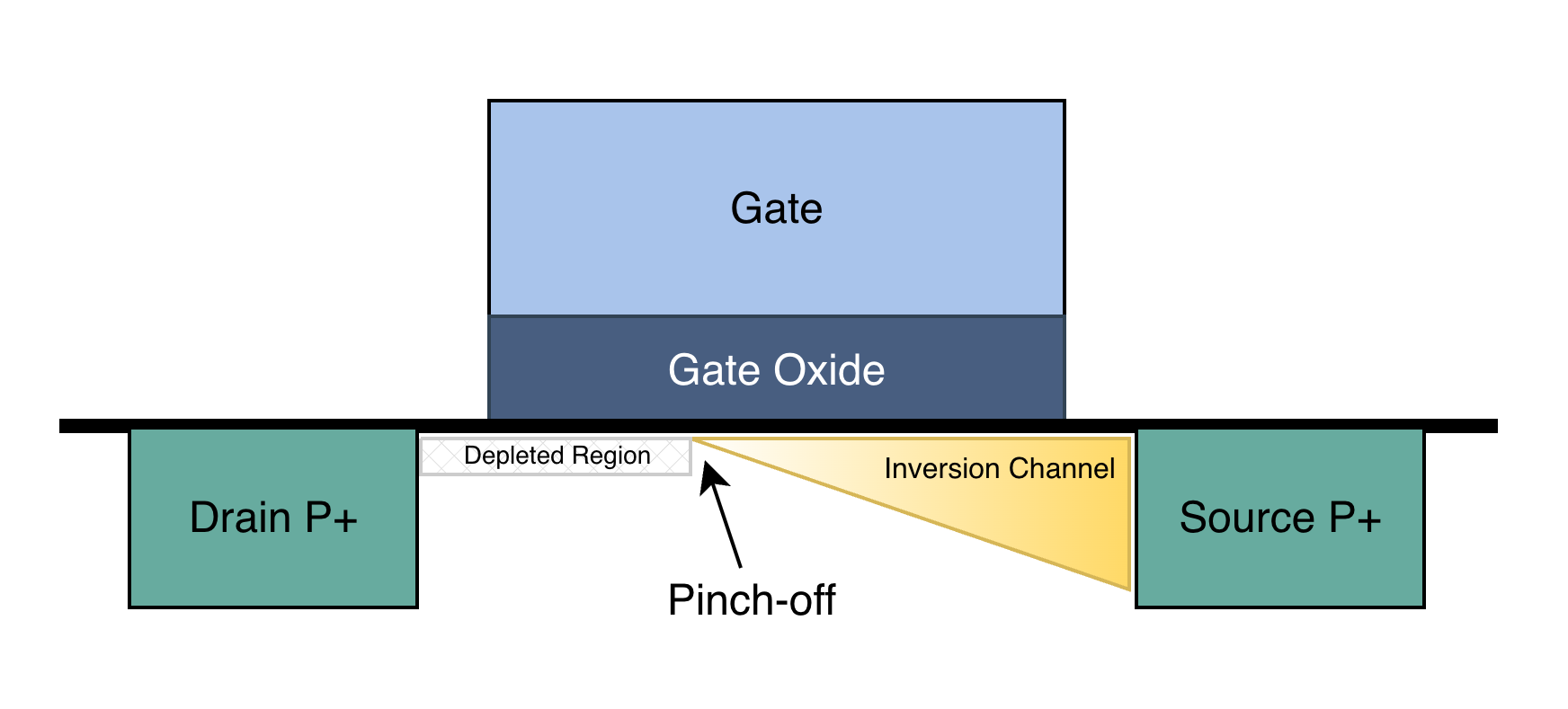}}
\caption{Impact ionization near the MOSFET drain terminal (figure not to scale). When the source/gate voltage exceeds a threshold voltage, an inversion layer forms and current begins to flow. Inversion is strongest near the source and will pinch off at a certain distance from the source implant. After the pinch-off point, the excess potential will create a strong electric field which draws the carriers to the drain implant. In this region, the impact ionization probability is high.}
\label{fig:impact_ionization_diagram}  
\end{figure}

Impact ionization may take place if a charge carrier in the strong electric field near the MOSFET drain implant gains enough kinetic energy to liberate an electron from an atom in the silicon substrate, creating an electron-hole pair. The liberated electrons, when collected at the internal gate (at a positive potential with respect to the channel), introduce excess noise (Fig. \ref{fig:impact_ionization_diagram}). 

To determine the probability of impact ionization, we calculate the average strength of the electric field between the pinch-off location and the drain using the known length of the MOSFET channel, the operating voltages of the source, drain, and gate, and a nominal threshold voltage. Because the electric field near the drain is highly nonlinear, it is not possible to calculate this value accurately using a simple linear model. Therefore, we include a correction factor, $K$, in our calculation to scale the electric field near the drain and investigate the effect on impact ionization probability. 

Using the coefficients from the the Van Overstraeten$\mbox{-}$De Man model\cite{Rivera23,VanOverstraeten70}, we calculate the number of impact collisions per charge carrier per unit length near the drain, 

\begin{equation}
\alpha = A(T) \exp{[-B(T)/KE]}
\end{equation}
where $A$ and $B$ are Overstraeten impact collision coefficients and are a function of temperature and the type carrier (hole in this case). $E$ is the electric field near the drain and $K$ is the correction factor in the electric field calculation.

The probability of one collision in length $L$ per carrier is given by 
\begin{equation}
p = 1 - exp (-\alpha~L).    
\end{equation}

If $N_{carr}$ is the number of carriers in time $t$, the number of electrons generated is
\begin{equation}
    N  = [ 1 - \exp{(-\alpha~L)}]~N_{carr} , 
\end{equation}
where $N_{carr}$ is given by
\begin{equation}
    N_{carr} = (I_{drain} / Q_e)~t .
\end{equation}

\begin{figure} [t!]
\centering
\begin{subfigure}{0.32\textwidth}    {\includegraphics[width=\textwidth]{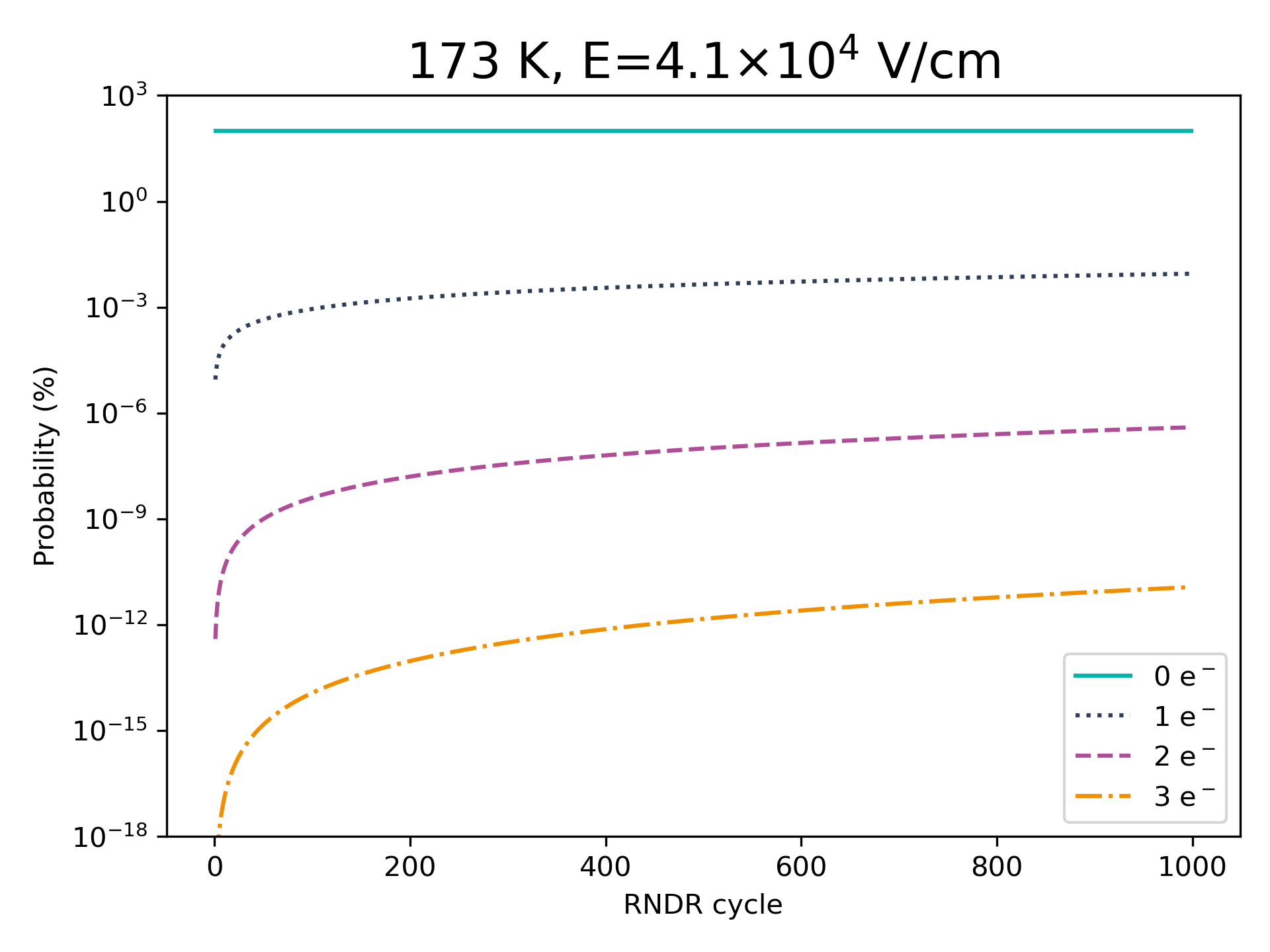}}
\end{subfigure}
\begin{subfigure}{0.32\textwidth}
    \includegraphics[width=\textwidth]{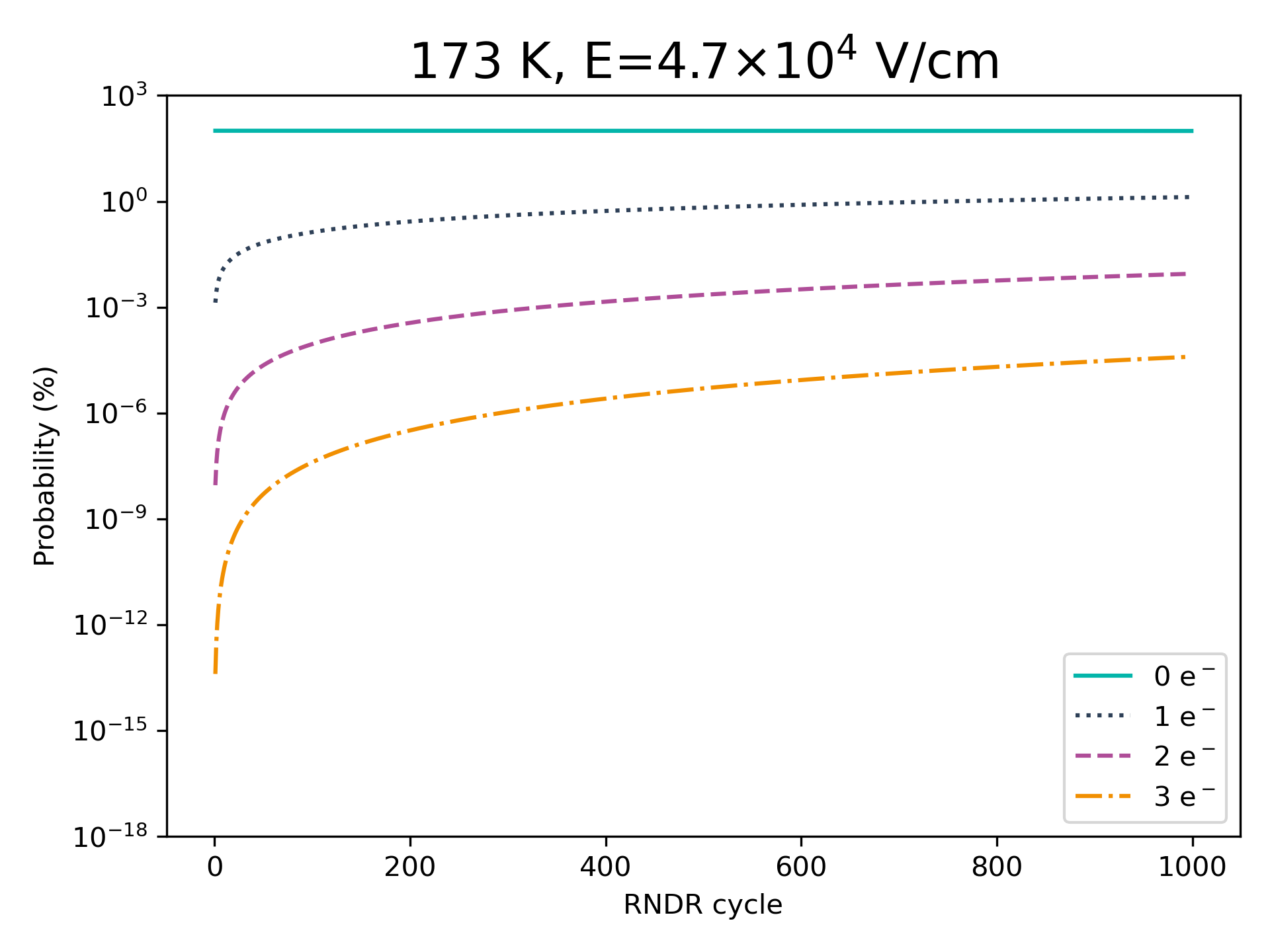}
\end{subfigure}
\begin{subfigure}{0.32\textwidth}
    \includegraphics[width=\textwidth]{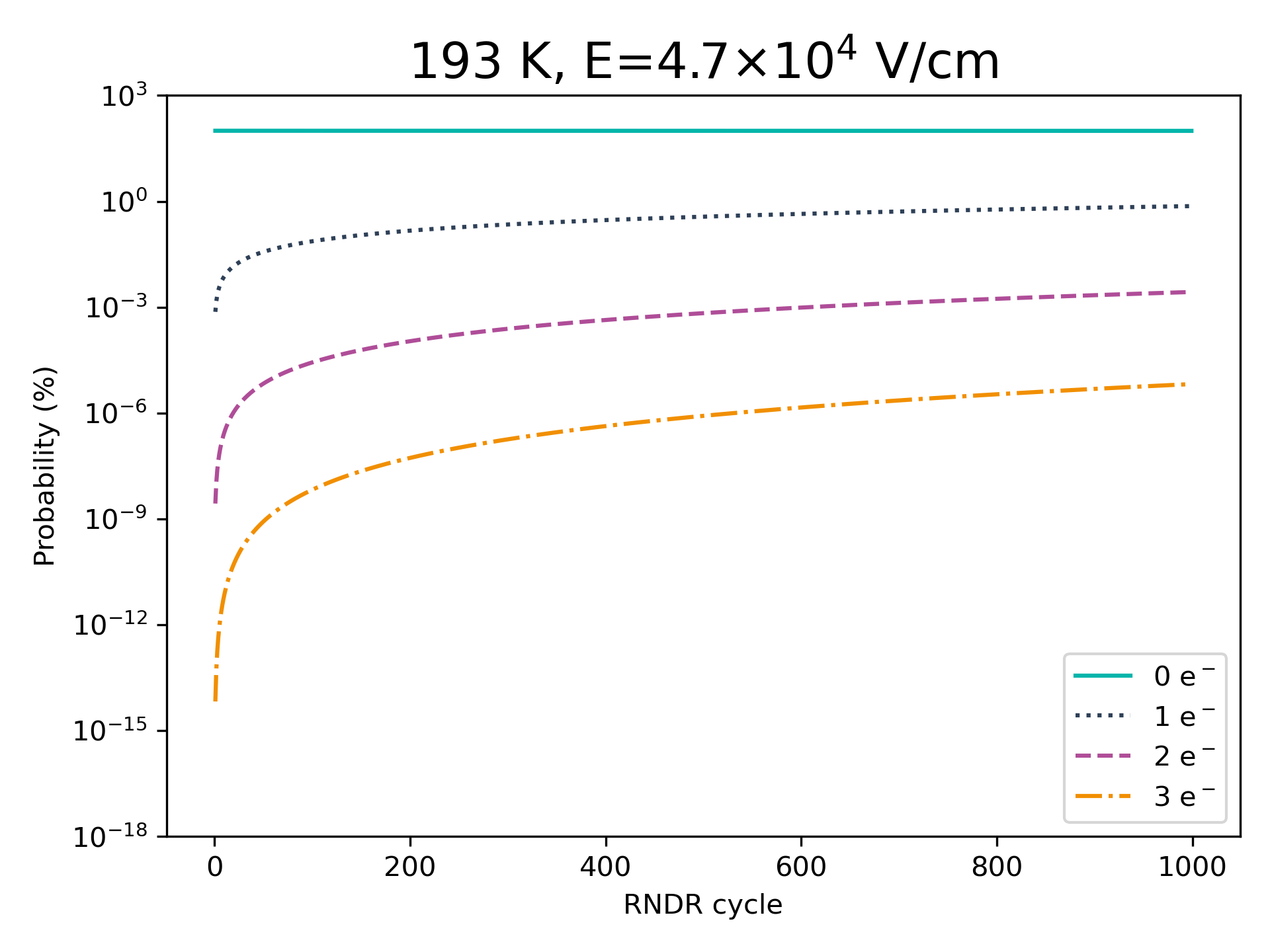}
\end{subfigure}
\caption{The probability of generating 0, 1, 2 or 3 electrons via impact ionization. The probability increases with the strength of the electric field near the drain and is inversely proportional to temperature.}
\label{fig:impact_prob_percent_model}  
\end{figure}

We then calculate the probability of generating $n$ electrons as a function of time per cycle,
\begin{equation}
\label{eq:prob_impact}
P(n)= \frac{N^n exp(-N)}{n!}\,.
\end{equation}
We find that the probability increases with the strength of electric field near the drain (Fig \ref{fig:impact_prob_percent_model}). We also find that the effect is expected to decrease at higher temperatures, likely due to the increase in lattice vibration or phonon excitation causing more scattering and preventing the charge carriers from gaining sufficient kinetic energy. 

\begin{figure} [b!]
\centering
\begin{subfigure}{0.49\textwidth}    {\includegraphics[width=\textwidth]{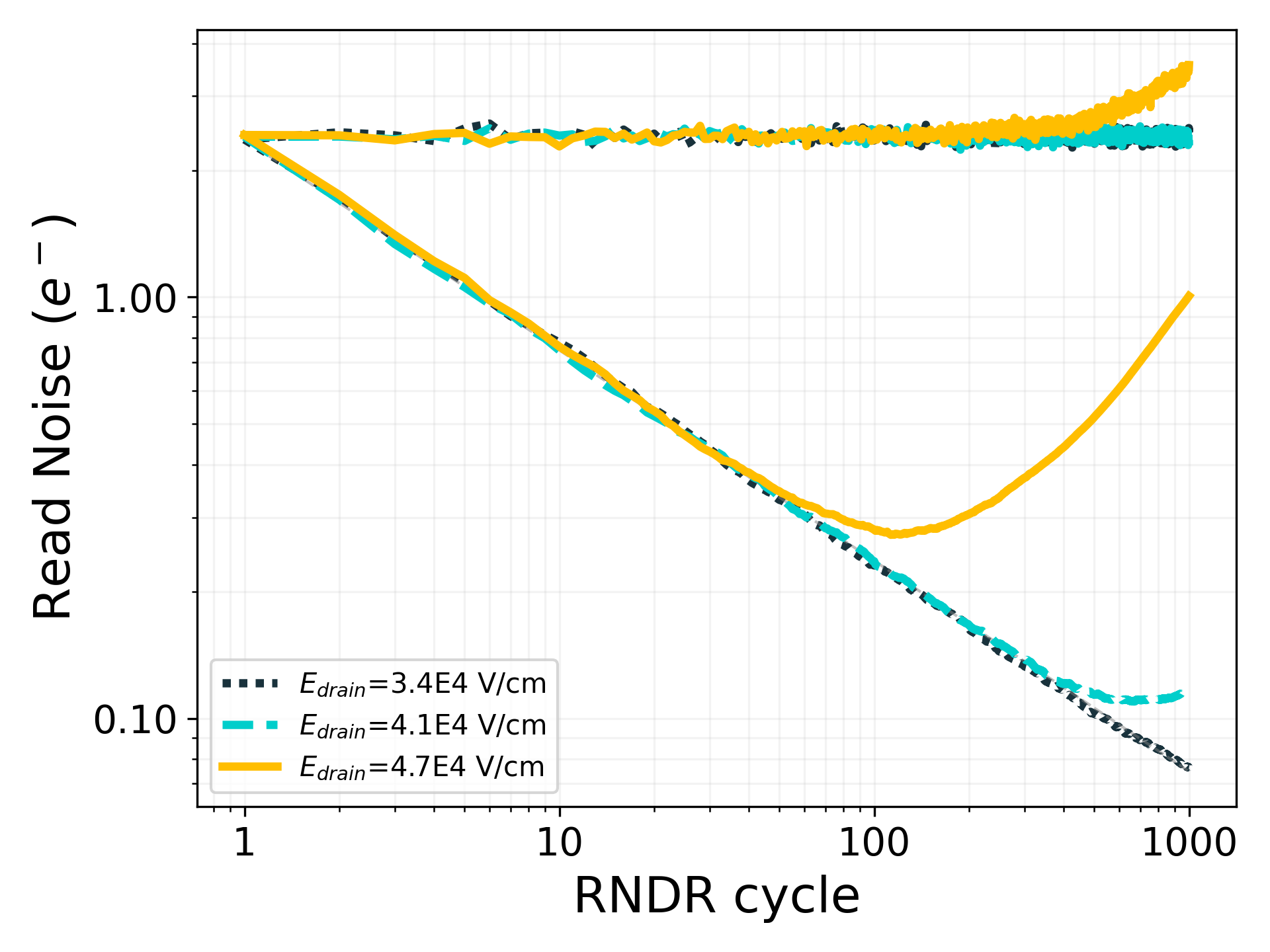}}
\end{subfigure}
\begin{subfigure}{0.49\textwidth}
    \includegraphics[width=\textwidth]{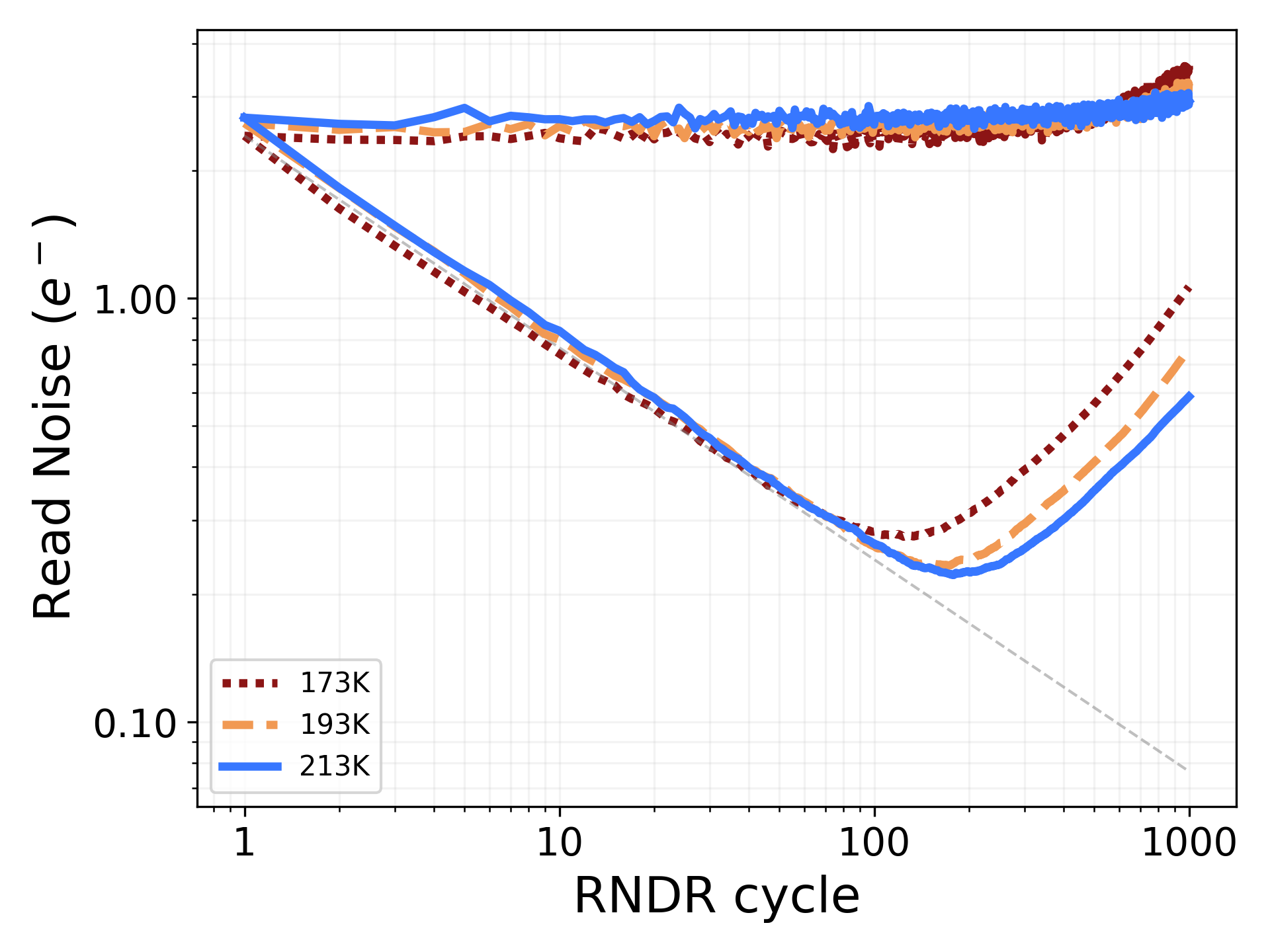}
\end{subfigure}
\caption{Simulation of the hypothetical situation in which we see non-negligible effects of impact ionization across 1000 RNDR cycles. Left: The impact ionization model at 173 K demonstrates that the effects are highly dependent on the strength of the electric field near the MOSFET drain. Such high electric field value causing excess noise can be ruled out in our devices based on the fact that we do not presently see evidence of excess noise in laboratory data up to 200 cycles\cite{PanJATIS2026}. 
Right:  Impact ionization model for three temperatures for $4.74\times10^{4}$ V/cm. 
}
\label{fig:impact_ionization_energy_temp}  
\end{figure}
We inject this probability into our Monte Carlo model and compare the predicted performance with varying electric fields (Fig. \ref{fig:impact_ionization_energy_temp} left) for a detector temperature of 173 K. We then explore the influence of the device temperature on the noise for an electric field of $4.74\times10^{4}$ V/cm (Fig. \ref{fig:impact_ionization_energy_temp} right).
Because we do not observe any deviation in read noise from the $1/\mathrm{\sqrt{N}}$ trend within our 200 RNDR cycle experiment,  we can rule out such high electric field causing excess noise due to impact ionization in our device.


\subsection{Upper Constraints on the SiSeRO noise floor}
\label{sec:upper_limit}  

Although no statistically significant excess charge is currently detected in our data, it is important to understand how many cycles it may take for an effect to be observed. Using the independent noise measurements at every RNDR cycle from our data, we can place an upper limit on the amplitude of the excess charge which may appear at longer RNDR cycles. 

Using laboratory measurements of first-generation SiSeROs operated at 100-200 RNDR cycles, we modeled the RNDR behavior as a constant and the total expected excess charge signal in the output as a linear increase in noise over time. We use $\chi^2$-minimization to fit a model, then performed an F-test for nested models in each dataset, comparing the full model to a model without an excess charge component. We find that the excess charge component does not provide a better fit to our data. We bootstrap each dataset to parametrize the error on the slope of the linear term and find that it is consistent with zero, again indicating there is no evidence of excess noise. Any excess charge produced is sufficiently small that it does not measurably affect detector performance under the operating conditions tested. 

\begin{figure} [htb]
\centering 
    {\includegraphics[width=0.5\textwidth]{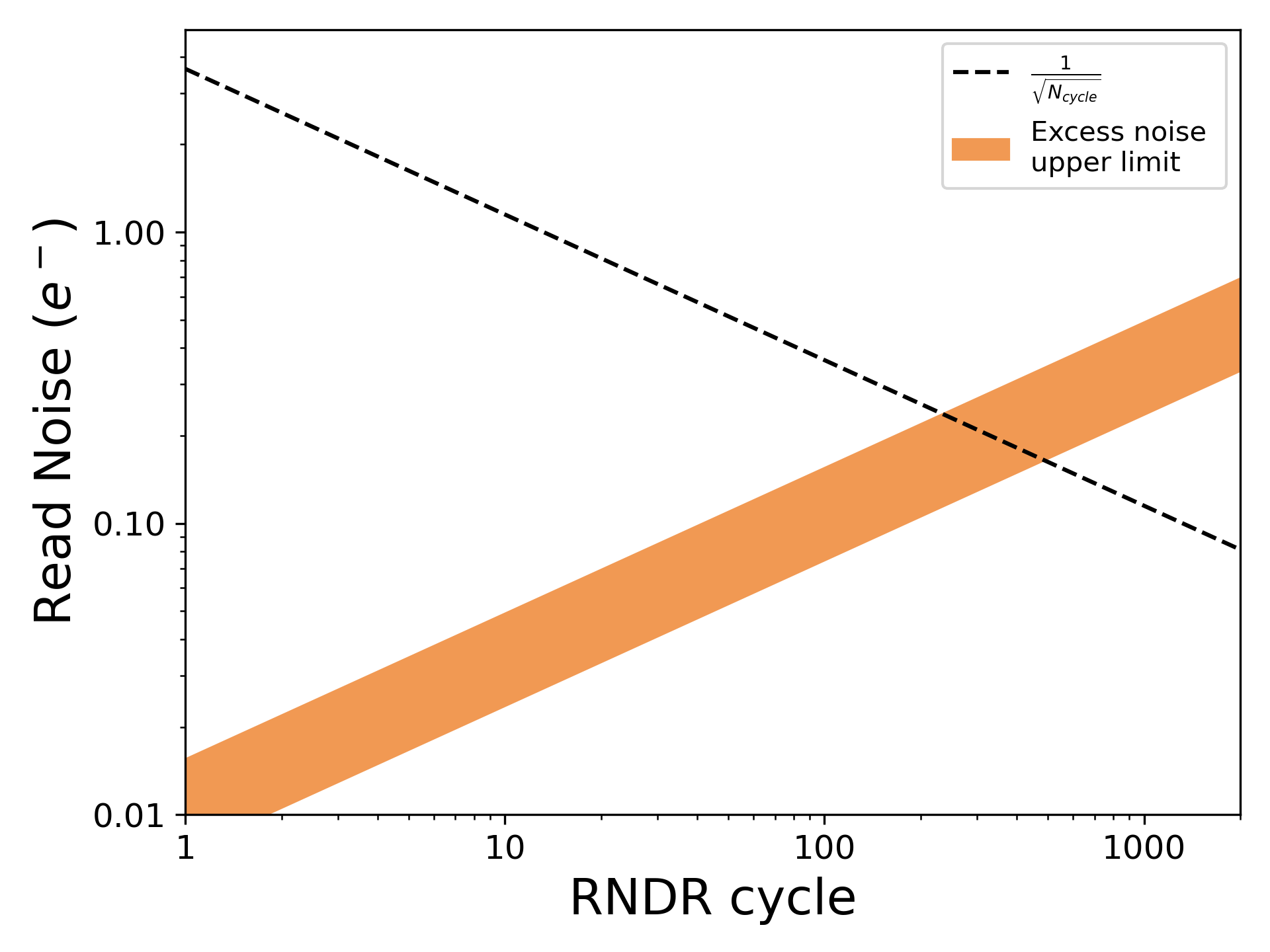}}
\caption{Statistical limits on the expected excess charge signal for 173 K: 2$\sigma$ upper bound of $3.08\times10^{-4}$ ADU/cycle (9.61 e$^-$/pixel/s), and a minimum signal of $7.23\times10^{-5}$ ADU/cycle (2.26 e$^-$/pixel/s)\cite{PanJATIS2026}.
}
\label{fig:upper_limit}  
\end{figure}

 Using this model, we place statistical constraints on the maximum dark current-like noise at various operating conditions (Fig. \ref{fig:upper_limit}). These upper limits on the amplitude of excess charge provide a constraint on detector performance and demonstrate that any dark current-like charge generated must lie below the derived limits. A detailed discussion of the upper bound calculations can be found in Pan et al. 2026 (in prep)\cite{PanJATIS2026}. 

\section{Summary}
\label{sec:summary}  

SiSeRO technology development provides a promising pathway to advance the strategic goals of the astronomy community by delivering fast, low-noise detectors with low dark current for future flagship telescopes in the X-ray and UV/visible/NIR regimes. In first-generation SiSeRO detectors, experimental data shows no deviation from the $1/\mathrm{\sqrt{N}}$ read noise trend up to maximum tests of 200 RNDR cycles (Pan et al. 2026, in prep)\cite{PanJATIS2026}. There is no hint of excess noise from known physical mechanisms, an encouraging result given that these first-generation SiSeRO prototypes were not optimized for RNDR readout. 

As we increase RNDR cycles to meet the noise requirements of future flagship space- and ground-based telescopes, we may eventually expect to see excess noise from various physical processes. To prepare for this, we have calculated the probability of electron emission as function of temperature, readout time, and the physical properties of the output stage, then probabilistically added it to an RNDR readout model. We have modeled the effects of trap-assisted thermal leakage and impact ionization in our detectors up to 1000 cycles and placed statistical constraints on these effects based on our laboratory measurements. 

Future work will entail laboratory testing of newly fabricated second-generation SiSeROs at significantly longer RNDR cycle runs.{\cite{pan2026_93pp}} If we observe any dark current-like signal, we will assess whether those terms are important in the context of future missions and accordingly adjust the design iterations of future SiSeRO devices.


\acknowledgments
\label{sec:acknowledgements}  
This work has been supported by the NASA Astrophysics Research and Analysis (APRA) program under grants 80NSSC22K1921,  80NSSC25K7957 and 80NSSC19K0499, as well as the NASA Strategic Astrophysics Technology (SAT) program under grant 80NSSC23K0211.


\bibliography{references} 

@unpublished{PanJATIS2026,
    author = {Abigail Y. Pan and Tanmoy Chattopadhyay and Tonya L. Peshel and Haley R. Stueber and Sven Herrmann and Peter Orel and Kevan Donlon and Steven W. Allen and Marshall W. Bautz and Michael Cooper and Beverly LaMarr and Chris Leitz and Eric Miller and R. Glenn Morris and Declan O'Neill and Artem Poliszczuk and Gregory Prigozhin and Ilya Prigozhin and Daniel R. Wilkins},
    title = {Demonstrating sub-electron noise performance in Single electron Sensitive Readout (SiSeRO) devices},
    note = {In preparation},
    year         = {2026},
}

@misc{HabWorldsRoadmap2024,
  author       = {Matthew R. Bolcar and Feng Zhao},
  title        = {Habitable Worlds Observatory Technology Roadmap},
  howpublished = {[PowerPoint slides]},
  year         = {2024},
  note         = {Exoplanet Exploration Program Office Webinar Series},
  url          = {https://ntrs.nasa.gov/api/citations/20240014303/downloads/ExEP%20HWO%20TechWebinar_20241114_v2.pdf}
}

@article{chattopadhyayImprovedNoisePerformance2023,
	title = {Improved noise performance from the next-generation buried-channel p-{MOSFET} {SiSeROs}},
	volume = {9},
	issn = {2329-4124, 2329-4221},
	url = {https://www.spiedigitallibrary.org/journals/Journal-of-Astronomical-Telescopes-Instruments-and-Systems/volume-9/issue-2/026001/Improved-noise-performance-from-the-next-generation-buried-channel-p/10.1117/1.JATIS.9.2.026001.full},
	doi = {10.1117/1.JATIS.9.2.026001},
	number = {2},
	urldate = {2025-07-02},
	journal = {Journal of Astronomical Telescopes, Instruments, and Systems},
	publisher = {SPIE},
	author = {Chattopadhyay, Tanmoy and Herrmann, Sven C. and Kaplan, Matthew and Orel, Peter and Donlon, Kevan and Prigozhin, Gregory Y. and Morris, Glenn R. and Cooper, Michael J. and Malonis, Andrew C. and Allen, Steven W. and Bautz, Marshall W. and Leitz, Chris W.},
	month = may,
	year = {2023},
	pages = {026001},
}

@inproceedings{chattopadhyayDemonstratingSubelectronNoise2024,
	title = {Demonstrating sub-electron noise performance in single electron sensitive readout ({SiSeRO}) devices},
	volume = {13103},
	url = {https://www.spiedigitallibrary.org/conference-proceedings-of-spie/13103/1310312/Demonstrating-sub-electron-noise-performance-in-single-electron-sensitive-readout/10.1117/12.3020855.full},
	doi = {10.1117/12.3020855},
	urldate = {2024-10-04},
	booktitle = {X-{Ray}, {Optical}, and {Infrared} {Detectors} for {Astronomy} {XI}},
	publisher = {SPIE},
	author = {Chattopadhyay, Tanmoy and Herrmann, Sven and Orel, Peter and Donlon, Kevan and Allen, Steven W. and Bautz, Marshall W. and Cantrall, Brianna J. and Cooper, Michael and LaMarr, Beverly and Leitz, Christopher and Miller, Eric and Morris, Glenn R. and Pan, Abigail Y. and Prigozhin, Gregory and Prigozhin, Ilya and Stueber, Haley and Wilkins, Daniel R.},
	month = aug,
	year = {2024},
	pages = {451--459},
}

@inproceedings{panDesignDevelopmentCommissioning2025,
	title = {Design, development, and commissioning of a flexible test setup for the {AXIS} prototype detector},
	volume = {13625},
	url = {https://www.spiedigitallibrary.org/conference-proceedings-of-spie/13625/136251J/Design-development-and-commissioning-of-a-flexible-test-setup-for/10.1117/12.3063468},
	doi = {10.1117/12.3063468},
	language = {en},
	urldate = {2026-06-23},
	booktitle = {{UV}, {X}-{Ray}, and {Gamma}-{Ray} {Space} {Instrumentation} for {Astronomy} {XXIV}},
	publisher = {SPIE},
	author = {Pan, Abigail Y. and Stueber, Haley R. and Chattopadhyay, Tanmoy and Allen, Steven W. and Bautz, Marshall W. and Donlon, Kevan and Grant, Catherine E. and Herrmann, Sven and LaMarr, Beverly and Malonis, Andrew and Miller, Eric D. and Morris, Glenn and Orel, Peter and Poliszczuk, Artem and Prigozhin, Gregory and Wilkins, Dan},
	month = sep,
	year = {2025},
	pages = {440},
}

@inproceedings{pan2026_93pp,
author = {Abigail Y. Pan and Declan O'Neill  and Kevan Donlon and Peter Orel and Sven Herrmann and Steven Allen and Marshall W. Bautz and Tanmoy Chattopadhyay and Michael Cooper and Catherine E. Grant and Jill Juneau and Chris Leitz and Beverly LaMarr and Eric D. Miller and Glenn Morris and Tonya L. Peshel and Artem Poliszczuk and Gregory Prigozhin and Ilya Prigozhin and Keith Warner and Haley R. Stueber},
title = {{First results for second generation SiSeRO CCD devices}},
volume = {14157-41},
booktitle = {X-Ray, Optical, and Infrared Detectors for Astronomy XII},
editor = {},
organization = {International Society for Optics and Photonics},
publisher = {SPIE},
pages = {},
year = {2026},
doi = {},
URL = {}
}

@book{nationalacademiesofsciencesPathwaysDiscoveryAstronomy2023,
	address = {Washington, DC},
	title = {Pathways to {Discovery} in {Astronomy} and {Astrophysics} for the 2020s},
	isbn = {978-0-309-46734-6},
	url = {https://nap.nationalacademies.org/catalog/26141/pathways-to-discovery-in-astronomy-and-astrophysics-for-the-2020s},
	doi = {10.17226/26141},
	publisher = {The National Academies Press},
	author = {National Academies of Sciences, {Engineering,}  and {Medicine}},
	year = {2023}
}

@ARTICLE{van_fet_noise_1962,
  author={Der Ziel, A. Van},
  journal={Proceedings of the IRE}, 
  title={Thermal Noise in Field-Effect Transistors}, 
  year={1962},
  volume={50},
  number={8},
  pages={1808-1812},
  doi={10.1109/JRPROC.1962.288221}
  }

@ARTICLE{Pullia04,
  author={Pullia, A. and Riboldi, S.},
  journal={IEEE Transactions on Nuclear Science}, 
  title={Time-domain Simulation of electronic noises}, 
  year={2004},
  volume={51},
  number={4},
  pages={1817-1823},
  doi={10.1109/TNS.2004.832564}
  }

@ARTICLE{Sah57,
  author={Sah, Chih-tang and Noyce, Robert N. and Shockley, William},
  journal={Proceedings of the IRE}, 
  title={Carrier Generation and Recombination in P-N Junctions and P-N Junction Characteristics}, 
  year={1957},
  volume={45},
  number={9},
  pages={1228-1243},
  doi={10.1109/JRPROC.1957.278528}
  }

@article{Shockley52,
  title = {Statistics of the Recombinations of Holes and Electrons},
  author = {Shockley, W. and Read, W. T.},
  journal = {Phys. Rev.},
  volume = {87},
  issue = {5},
  pages = {835--842},
  numpages = {0},
  year = {1952},
  month = {Sep},
  publisher = {American Physical Society},
  doi = {10.1103/PhysRev.87.835},
  url = {https://link.aps.org/doi/10.1103/PhysRev.87.835}
}

@ARTICLE{Rivera23,
  author={Currás Rivera, Esteban and Moll, Michael},
  journal={IEEE Transactions on Electron Devices}, 
  title={Study of Impact Ionization Coefficients in Silicon With Low Gain Avalanche Diodes}, 
  year={2023},
  volume={70},
  number={6},
  pages={2919-2926},
  doi={10.1109/TED.2023.3267058}
  }

@article{VanOverstraeten70,
    author = {Van Overstraeten, R. and De Man, H.},
    title = {Measurement of the ionization rates in diffused silicon p-n junctions},
    journal = {Solid-State Electronics},
    volume={13},
    issue={5},
    pages={583-608},
    year ={1970},
    doi={10.1016/0038-1101(70)90139-5}
}

@article{Stefanov2020_skippercmos,
  title={Simulations and Design of a Single-Photon CMOS Imaging Pixel Using Multiple Non-Destructive Signal Sampling},
  author={Konstantin D. Stefanov and Martin J. Prest and Mark Downing and Elizabeth M. George and Naidu Bezawada and Andrew D. Holland},
  journal={Sensors (Basel, Switzerland)},
  year={2020},
  volume={20},
  url={https://api.semanticscholar.org/CorpusID:214802950}
}

@inproceedings{herrmann20_mcrc,
author = {Sven Herrmann and Josephine Wong and Tanmoy Chattopadhyay and R. Glenn Morris and Barry Burke and Gregory Prigozhin and Mike Cooper and David Craig and Kevan Donlon and Richard Foster and Andrew Malonis and Mark Bautz and Steve Allen},
title = {{MCRC V1: development of integrated readout electronics for next generation x-ray CCD detectors for future satellite observatories}},
volume = {11454},
booktitle = {X-Ray, Optical, and Infrared Detectors for Astronomy IX},
editor = {Andrew D. Holland and James Beletic},
organization = {International Society for Optics and Photonics},
publisher = {SPIE},
pages = {412 -- 418},
year = {2020},
doi = {10.1117/12.2561741},
URL = {https://doi.org/10.1117/12.2561741}
}

@inproceedings{bautz20,
author = {M. Bautz and B. Burke and M. Cooper and D. Craig and K. Donlon and R. Foster and C. E. Grant and B. LaMarr and C. Leitz and A. Malonis and E. Miller and G. Prigozhin and C. Thayer and S. Allen and S. Herrmann and T. Chattopadhyay and R. G. Morris},
title = {{Progress toward fast, low-noise, low-power CCDs for Lynx and other high-energy astrophysics missions}},
volume = {11444},
booktitle = {Space Telescopes and Instrumentation 2020: Ultraviolet to Gamma Ray},
editor = {Jan-Willem A. den Herder and Shouleh Nikzad and Kazuhiro Nakazawa},
organization = {International Society for Optics and Photonics},
publisher = {SPIE},
pages = {1318 -- 1323},
year = {2020},
doi = {10.1117/12.2561441},
URL = {https://doi.org/10.1117/12.2561441}
}

@article{chattopadhyay22_sisero,
author = {Tanmoy Chattopadhyay and Sven Herrmann and Barry E. Burke and Kevan Donlon and Gregory Prigozhin and Glenn Morris and Peter Orel and Michael Cooper and Andrew Malonis and Daniel R. Wilkins and Vyshnavi Suntharalingam and Steven W. Allen and Marshall W. Bautz and Chris Leitz},
title = {{First results on SiSeRO devices: a new x-ray detector for scientific instrumentation}},
volume = {8},
journal = {Journal of Astronomical Telescopes, Instruments, and Systems},
number = {2},
publisher = {SPIE},
pages = {1 -- 12},
year = {2022},
doi = {10.1117/1.JATIS.8.2.026006},
URL = {https://doi.org/10.1117/1.JATIS.8.2.026006}
}

@article{chattopadhyay22_ccd,
author = {Tanmoy Chattopadhyay and Sven C. Herrmann and Peter Orel and R. Glenn Morris and Gregory Y. Prigozhin and Andrew C. Malonis and Richard F. Foster and David M. Craig and Barry E. Burke and Steven W. Allen and Marshall W. Bautz},
title = {{Development and characterization of a fast and low noise readout for the next generation x-ray charge-coupled devices}},
volume = {8},
journal = {Journal of Astronomical Telescopes, Instruments, and Systems},
number = {2},
publisher = {SPIE},
pages = {1 -- 12},
year = {2022},
doi = {10.1117/1.JATIS.8.2.026005},
URL = {https://doi.org/10.1117/1.JATIS.8.2.026005}
}

@inproceedings{Bautzetal2022,
	author = {{Bautz}, M.~W. and {Foster}, R. and {Grant}, C.~E. and {LaMarr}, B. and {Malonis}, A. and {Miller}, E.~D. and {Prigozhin}, G. and {Burke}, B. and {Cooper}, M. and {Donlon}, K. and {Lambert}, R. and {Warner}, K. and {Young}, D. and {Chattopadhyay}, T. and {Herrmann}, S. and {Morris}, R.~G. and {Leitz}, C. and {Allen}, S.},
	booktitle = {Space Telescopes and Instrumentation 2022: Ultraviolet to Gamma Ray},
	pages = {12181-85},
	series = {Society of Photo-Optical Instrumentation Engineers (SPIE) Conference Series},
	title = {{Performance of high frame-rate CCDs for future strategic missions}},
	volume = {12181},
	year = 2022}

@article{wolfel06,
title = {Sub-electron noise measurements on repetitive non-destructive readout devices},
journal = {Nuclear Instruments and Methods in Physics Research Section A: Accelerators, Spectrometers, Detectors and Associated Equipment},
volume = {566},
number = {2},
pages = {536-539},
year = {2006},
issn = {0168-9002},
doi = {https://doi.org/10.1016/j.nima.2006.06.060},
url = {https://www.sciencedirect.com/science/article/pii/S0168900206012113},
author = {Stefan Wölfel and Sven Herrmann and Peter Lechner and Gerhard Lutz and Matteo Porro and Rainer Richter and Lothar Strüder and Johannes Treis}
}

@ARTICLE{rodrigues21,
       author = {{Rodrigues}, Dario and {Andersson}, Kevin and {Cababie}, Mariano and {Donadon}, Andre and {Botti}, Ana and {Cancelo}, Gustavo and {Estrada}, Juan and {Fernandez-Moroni}, Guillermo and {Piegaia}, Ricardo and {Senger}, Matias and {Haro}, Miguel Sofo and {Stefanazzi}, Leandro and {Tiffenberg}, Javier and {Uemura}, Sho},
        title = "{Absolute measurement of the Fano factor using a Skipper-CCD}",
      journal = {Nuclear Instruments and Methods in Physics Research A},
         year = 2021,
        month = sep,
       volume = {1010},
          eid = {165511},
        pages = {165511},
          doi = {10.1016/j.nima.2021.165511},
archivePrefix = {arXiv},
       eprint = {2004.11499},
 primaryClass = {physics.ins-det},
       adsurl = {https://ui.adsabs.harvard.edu/abs/2021NIMPA101065511R}
}

@article{tiffenberg17,
  title = {Single-Electron and Single-Photon Sensitivity with a Silicon Skipper CCD},
  author = {Tiffenberg, Javier and Sofo-Haro, Miguel and Drlica-Wagner, Alex and Essig, Rouven and Guardincerri, Yann and Holland, Steve and Volansky, Tomer and Yu, Tien-Tien},
  journal = {Phys. Rev. Lett.},
  volume = {119},
  issue = {13},
  pages = {131802},
  numpages = {6},
  year = {2017},
  month = {Sep},
  publisher = {American Physical Society},
  doi = {10.1103/PhysRevLett.119.131802},
  url = {https://link.aps.org/doi/10.1103/PhysRevLett.119.131802}
}

@inproceedings{treberspurg22_rndrdepfet,
author = {W. Treberspurg and A. B{\"a}hr and H. Kluck and P. Lechner and J. Ninkovic and J. Treis and H. Shi and J. Schieck},
title = {{Performance of a kilo-pixel RNDR-DEPFET detector}},
volume = {12191},
booktitle = {X-Ray, Optical, and Infrared Detectors for Astronomy X},
editor = {Andrew D. Holland and James Beletic},
organization = {International Society for Optics and Photonics},
publisher = {SPIE},
pages = {1219119},
year = {2022},
doi = {10.1117/12.2629248},
URL = {https://doi.org/10.1117/12.2629248}
}

@article{chattopadhyay24_rnrdr,
author = {Tanmoy Chattopadhyay and Sven Herrmann and Peter Orel and Kevan Donlon and Gregory Prigozhin and Glenn Morris and Michael Cooper and Beverly LaMarr and Andrew Malonis and Steven W. Allen and Marshall W. Bautz and Chris Leitz},
title = {{Demonstrating repetitive non-destructive readout with SiSeRO devices}},
volume = {10},
journal = {Journal of Astronomical Telescopes, Instruments, and Systems},
number = {1},
publisher = {SPIE},
pages = {016004},
year = {2024},
doi = {10.1117/1.JATIS.10.1.016004},
URL = {https://doi.org/10.1117/1.JATIS.10.1.016004}
}

@inproceedings{stueber2024,
        author = {Haley R. Stueber and Tanmoy Chattopadhyaya and Sven C. Herrmann and Peter Orel and Tsion
Gebre and Aanand Joshi and Steven W. Allen and Glenn Morris and Artem Poliszczuk},
        title = {{The XOC X-ray Beamline: Probing Colder, Quieter, and Softer}},
        volume = {13103},
        booktitle = {X-Ray, Optical, and Infrared Detectors for Astronomy XI},
        editor = {},
        organization = {International Society for Optics and Photonics},
        publisher = {SPIE},
        pages = {13103-77},
        year = {2024},
        doi = {},
        URL = {}
}

@inproceedings{herrmann2024,
        author = {Sven C. Herrmann and Peter Orel and Tanmoy Chattopadhyay and Glenn R. Morris and Gregory Y. Prigozhin and Haley R. Stueber and Steven W. Allen and Marshall W. Bautz and Kevan Donlon and Beverly J. LaMarr and Christopher W. Leitz and Eric D. Miller and Artem Poliszczuk and Dan R. Wilkins},
        title = {{Continued developments in X-ray speed reading: fast, low noise readout for next-generation wide-field imagers}},
        volume = {13103},
        booktitle = {X-Ray, Optical, and Infrared Detectors for Astronomy XI},
        editor = {},
        organization = {International Society for Optics and Photonics},
        publisher = {SPIE},
        pages = {1310383},
        year = {2024},
        doi = {},
        URL = {}
}

@article{Lapi2024_skipperMAS,
author = {Agustin J. Lapi and Blas J. Irigoyen Gimenez and Miqueas E. Gamero and Claudio R. Chavez Blanco and Fernando Chierchie and Guillermo Fernandez-Moroni and Stephen Holland and Ana M. Botti and Brenda A. Cervantes-Vergara and Javier Tiffenberg and Juan Estrada},
title = {{Sixteen multiple-amplifier sensing charge-coupled devices and characterization techniques targeting the next generation of astronomical instruments}},
volume = {11},
journal = {Journal of Astronomical Telescopes, Instruments, and Systems},
number = {1},
publisher = {SPIE},
pages = {011203},
year = {2024},
doi = {10.1117/1.JATIS.11.1.011203},
URL = {https://doi.org/10.1117/1.JATIS.11.1.011203}
}

@ARTICLE{Lapi2024_skipperCMOS,
  author={Lapi, Agustin J. and Sofo-Haro, Miguel and Parpillon, Benjamin C. and Birman, Adi and Fernandez-Moroni, Guillermo and Rota, Lorenzo and Alcalde Bessia, Fabricio and Gupta, Aseem and Chavez Blanco, Claudio R. and Chierchie, Fernando and Segal, Julie and Kenney, Christopher J. and Dragone, Angelo and Li, Shaorui and Braga, Davide and Fenigstein, Amos and Estrada, Juan and Fahim, Farah},
  journal={IEEE Transactions on Electron Devices}, 
  title={Skipper-in-CMOS: Nondestructive Readout With Subelectron Noise Performance for Pixel Detectors}, 
  year={2024},
  volume={71},
  number={11},
  pages={6843-6849},
  doi={10.1109/TED.2024.3463631}}

@inproceedings{porelMCRCspie2024,
        author = {Peter Orel and Sven Herrmann and Tanmoy Chattopadhyay and Glenn R. Morris and Haley Stueber and Abby Pan and Steven W. Allen and Daniel Wilkins and Gregory Y. Prigozhin and Beverly LaMarr and Richard Foster and Andrew Malonis and Marshall W. Bautz and Michael J. Cooper and Kevan Donlon},
        title = {{X-ray speed reading with the MCRC: prototype success and next generation upgrades}},
        volume = {13103},
        booktitle = {X-Ray, Optical, and Infrared Detectors for Astronomy XI},
        editor = {},
        organization = {International Society for Optics and Photonics},
        publisher = {SPIE},
        pages = {1310332},
        year = {2024},
        doi = {},
        URL = {}
}

@inproceedings{stueber_ccd_2025,
author = {Haley R. Stueber and Abigail Y. Pan and Tanmoy Chattopadhyay and Steven W. Allen and Marshall W. Bautz and Kevan Donlon and Catherine E. Grant and Sven Herrmann and Beverly J. LaMarr and Andrew Malonis and Eric D. Miller and R. Glenn Morris and Peter Orel and Artem Poliszczuk and Gregory Y. Prigozhin and Daniel R. Wilkins},
title = {{Fast, low noise CCD systems for future strategic x-ray missions}},
volume = {13625},
booktitle = {UV, X-Ray, and Gamma-Ray Space Instrumentation for Astronomy XXIV},
editor = {Oswald H. Siegmund and Keri Hoadley},
organization = {International Society for Optics and Photonics},
publisher = {SPIE},
pages = {136251Q},
year = {2025},
doi = {10.1117/12.3064612},
URL = {https://doi.org/10.1117/12.3064612}
}

@inproceedings{bautz_ccd_2024,
author = {Marshall W. Bautz and Eric D. Miller and Gregory Y. Prigozhin and Beverly J. LaMarr and Andrew Malonis and Richard Foster and Catherine E. Grant and Benjamin Schneider and Christopher Leitz and Kevan Donlon and Ilya Prigozhin and Renee Lambert and Michael Cooper and Sven C. Herrmann and Peter Orel and Tanmoy Chattopadhyay and R. Glenn Morris and Daniel R. Wilkins and Haley R. Stueber and Artem Poliszczuk and Steven W. Allen},
title = {{Fast, low-noise image sensor technology for strategic x-ray astrophysics missions}},
volume = {13093},
booktitle = {Space Telescopes and Instrumentation 2024: Ultraviolet to Gamma Ray},
editor = {Jan-Willem A. den Herder and Shouleh Nikzad and Kazuhiro Nakazawa},
organization = {International Society for Optics and Photonics},
publisher = {SPIE},
pages = {130931Q},
year = {2024},
doi = {10.1117/12.3019193},
URL = {https://doi.org/10.1117/12.3019193}
}

@inproceedings{bredthauerArchonModernController2014,
	title = {Archon: {A} modern controller for high performance astronomical {CCDs}},
	volume = {9147},
	shorttitle = {Archon},
	url = {https://www.spiedigitallibrary.org/conference-proceedings-of-spie/9147/1/Archon-A-modern-controller-for-high-performance-astronomical-CCDs/10.1117/12.2058402},
	doi = {10.1117/12.2058402},
	language = {en},
	urldate = {2026-06-23},
	booktitle = {Ground-based and {Airborne} {Instrumentation} for {Astronomy} {V}},
	publisher = {SPIE},
	author = {Bredthauer, Greg},
	month = jul,
	year = {2014},
	pages = {1730},
}
\bibliographystyle{spiebib} 

\end{document}